# Nanosurgery: Observation of Peptidoglycan Strands in *Lactobacillus helveticus* Cell Walls

**Max Firtel,**

Department of Microbiology, University of Toronto, Ontario, M5S 1A8 Canada
(Deceased)

**Grant Henderson**

Department of Geology, University of Toronto, Ontario, M5S 1B3 Canada

**Igor Sokolov,**

Department of Physics, Clarkson University, Potsdam, NY, 13699-5820, USA

**Corresponding author**: Igor Sokolov, Clarkson Ave., Department of Physics,
Clarkson University, Potsdam, NY, 13699-5820, USA;
Tel. (315)268-2375
FAX: (315)268-6610
e-mail isokolov@clarkson.edu

ABSTRACT:

The internal cell wall structure of the bacterium *Lactobacillus helveticus* has been observed *in situ* in aqueous solution using an atomic force microscope (AFM). The AFM tip was used not only for imaging but presumably to remove mechanically large patches of the outer cell wall after appropriate chemical treatment, which typically leaves the bacteria alive. The surface exposed after such a "surgery" revealed ~26 nm thick twisted strands within the cell wall. The structure and location of the observed strands are consistent with the glycan backbone of peptidoglycan fibers that give strength to the cell wall. The found structural organization of these fibers has not been observed previously.

## 1. Introduction

Atomic force microscopy (AFM) has been successfully applied to the 3D *in situ* high-resolution visualization of soft biological surfaces (e.g., [1-3]) immersed in aqueous solutions. The lateral resolution in these studies can be on the order of nanometers while the vertical resolution depends on the elastic properties of the material under study, but can reach as high as parts of angstrom.  Here we report the use of the AFM with dual purpose.  First, we use it as a microscope to both observe the outer layers of the cell wall of the bacteria of *Lactobacillus helveticus*; and secondly, to induce some mechanical action of the AFM tip  onto the to observed bacterial surface. We hypothesize that the observed structure is presumably a result of "surgically" removed outer parts of the bacterial wall due to the mechanical action of the AFM tip. It should be noted that the mechanical action of the AFM tip was first reported in [4] to remove the top membrane from gap junctions isolated from rat liver cells. After removing, the scanning was believed to image the extracellular surface of the bottom membrane of the gap junction. In this paper, we demonstrate that the AFM tip can work as a "surgery" tool for studying the inner structure of biological surfaces *in-situ*. Specifically, we studied the inner structure of bacterial cell wall without its isolation, directly on the bacterial surface. Another major difference compared to the method reported in [4] is that we used some preliminary chemical treatment of the surface. (Such treatment does not kill the cells). The reported here "surgery" was not possible without the preliminary treatment, while the treatment used in [4] did not make any difference in the membrane removal. Using our method, we report a discovery of a new structure, presumably striations of peptidoglyken inside the cell wall.

In Gram-positive cells such as *L. helveticus*, *the* cell envelope consists of the cytoplasmic membrane, a 20-80 nm thick cell wall mainly composed of linear polymers of disaccharide pentapeptide subunits (i.e., peptidoglycan), and a protein crystal or S-layer. Fig. 1(a) shows a schematic cross-section of a Gram positive bacterium wall. More specifically, the wall includes anionic polymers such as teichoic acid which are cross-linked to the N-acetylmuramyl (NAM) and residue of the peptidoglycan, N-acetyl glucosamine (NAG); and lesser amounts of membrane bound lipoteichoic acid, neutral carbohydrates, and proteins (see, e.g., paper [5]). Peptidoglycan is composed of chains of peptidoglycan monomers (NAG-NAM-tetrapeptide). These monomers join together to form chains and the chains are then joined by cross-links between the tetrapeptides to provide strength, Fig.1 (b).

For *Bacillus subtilis,* the best characterized of all Gram-positive cells, the cell wall is described as a "gel" in which the stress-bearing glycan strands are underline, or embedded, in an amorphous anionic matrix of teichoic acid and other polymers which affect the charge density, porosity and to some extent the mechanical properties of the cell wall [6]. About ninety percent of the Gram-positive cell wall is comprised of peptidoglycan.

In the present work, the S-layer was removed by a chemical treatment of the bacteria, which were consequently immersed in broth for overnight. Finally, the exposed bacterial surface was mechanically treated with the AFM tip. Unusual striation reported in the present work was observed only after this sequence of treatments. Presumably, the dynamic action between the AFM tip and the cell wall was the essential component of the scanning procedure.

## 2. Materials and Methods

In the present study all the images were obtained using a Digital Instruments NanoScope IIIa Multimode AFM operating in contact mode under DI distilled water (pH *ca.* 7). The *A+B* signal of the feedback was about 3 *V* while the initial A-B signal was set at around -1 *V* and the setpoint at 0 *V*. The corresponding contact force was on the order of one to two nano-Newtons. Standard silicon nitride integrated pyramidal tips with an estimated radius of curvature of ∼50 nm were fixed on 200 $\mu m$ "narrow leg" cantilevers (spring constant $k$∼0.04±0.01 *N/m)*. The D scan head (maximum scan area is $12.5 \times 12.5 \ \mu m^2$, z-sensitivity is 9 nm/V) was employed throughout the study.

*L.helveticus* ATCC 12046 was maintained on MRS agar [10] incubated at 43°C in 5% $CO_2$ atm. Cells were collected by centrifugation (12,500 x g; 10 min), washed once in distilled water, and resuspended in 5 M LiCl (10 -15 mg of moist pellet mL$^{-1}$ LiCl) for 10 min at 0C, followed by washing in distilled water and centrifugation at 12,500 x g for 20 min. The purpose of the LiCl treatment is to remove the outer layer of the bacteria wall, which consists of S-protein and is called the S-layer. Such a treatment keeps the majority of bacteria alive (see, e.g., [3]). The cells are then placed back in the broth solution and allowed to settle on the surface of a sheet of mica overnight at 4°C. This has two effects: 1) it acts to bind the cells to the mica substrate and 2) the low temperature acts to inhibit S-layer regeneration [3,7]. Without these two effects we are unable to make the nanosurgery.

## 3. Results and Discussion

The second treatment, in MRS broth, at 4$^{o}$C overnight is not enough to regenerate the S-layer. Consequently, by the time the surface is scanned by AFM we are essentially scanning a cell wall that has little or no outer S-layer. On the other hand, it seems to enable the outer layers of the cell wall to become either "sticky", or possibly rigid. This improves the physical interaction between the AFM tip and the cell surface. We speculate that the second treatment in MRS broth somehow modifies the smooth surface of the bacteria left after the first treatment so that the AFM tip is better able to bind to the peptidoglycan layer. Consequently, the tip is able to physically "peel" off a part of the outside surface of the cell wall, thereby exposing the strand–like features observed within the cell wall, Figure 3a.

It should be noted that not every bacterium treated as described above result in the nanosurgery, and consequently, exhibited the striation. In about 50 measurements that were done in this study, we were able to observe the reported effect only in four cases. The most frequent images of the treated surface were similar to Fig.2, which is close to previously reported images (see Fig.2 in paper [3]), in which there were no striations observed. It is interesting to note that an attempt to use NaCl rather than LiCl in the treatment of the cells presumably leads to the nanosurgery but results in more poor images of the strands. The best image we obtained with NaCl treatment is shown in Figure 3b. Figure 3b resembles a typical scan in which we do not observe details of the internal cell structure. The observed images appear "washed out" in comparison with the LiCl treatment method. It should be noted that such a washy image is not easy to separate from an artifact. So, we cannot unambiguously claim that the nanosurgery was attained in the experiments with the NaCl treatment.

The image scans in which the peptidoglycan strands were observed (LiCl treatment) initially exhibited evidence of noise, typical of conditions in which weakly bound material is being removed from the surface. Scan contact can be lost at this moment. After the reengaging the AFM scanning, some residual noise, which corresponds to the surface fragments that have been removed, was eliminated after a few scans of the AFM tip across the cell surface. This indicates that the residual smooth surface had been removed. Fig.4 shows a high resolution of such a scan. One can see the cell wall fragments that are been removed, remain visible within the exposed area (indicated by the white arrows). The strands tended to be visible on the top of the cell (the region of maximum contact share pressure) which would lend credence to our assumption that the tip is physically removing a part of the smooth surface of the cell wall.

As one can see from Fig.3a and 4, the surface of the treated bacteria reveals ~26nm thick strands which are orientated around the circumference of the cylindrical shaped cell. Some of the strands seem to be twisted pairs as indicated by black arrows in Fig.5. The more stable components of the cell wall, such as peptidoglycan, undergo turnover over a period of generations [8]. The twisted appearance of the strands observed in this study are consistent with the helical arrangement proposed for glycan [9]. The thickness of the strands suggests that they are each composed of a bundle of glycan polymers. The strand lengths could not be determined since the ends of the strands continue beneath the smooth membrane. However, their length is at least 400 nm. It is naturally to expect that the strands observed surround entirely the bacterium.

## 4. Conclusions

We report the use of the AFM technique to study internal cell wall structure of the bacterium *Lactobacillus helveticus in-situ* while the bacteria were immersed in DI water of neutral pH. The AFM was utilized not only for imaging but presumably for mechanical removing large patches of the outer cell wall after appropriate chemical treatment which typically leaves the bacteria alive. The surface exposed after such a "surgery" revealed ~26 nm thick sometime twisted strands within the cell wall, which have not been observed previously. The structure and location of the observed strands are consistent with the glycan backbone of peptidoglycan fibers that give strength to the cell wall.

The presented work can open a new direction in study of structure of biological objects by means of the AFM technique.

**Acknowledgements**. This study was supported by NSERC research and equipment grants to GSH. P. Sivarajah is thanked for assistance with cell preparation and J. Campbell for support and encouragement. Our co-author and friend, Max Firtel, passed away suddenly prior to completion of this manuscript. His intellect, advice, wit, and friendship are sorely missed by us both.

List of Captions

**Figure 1.**  (a) Schematic structure of a Gram positive bacterial envelope.

(b) Structure of peptidoglycan in the cell wall. It is composed of chains of peptidoglycan monomers, N-acetyl glucosamine (NAG), N-acetylmuramyl (NAM), and tetrapeptide.

**Figure 2.** The surface of the bacterium *Lactobacillus helveticus* just after removal of the S-layer. Lateral scale of the image is 380 nm

**Figure 3.** Surface of bacterium after chemical treatment.

(a) LiCl treatment.  The lateral size of the image is 1900 nm.  The upper surface of the cell shows a set of striations some 20-30nm in width and which have not been observed before (enlarged view is shown in Fig.4).

(b) NaCl treatment. Lateral scale is 1300nm. The image is less clear relative to (a) above.

**Fig.4.** Enlarged view of the chemically treated bacterium after the AFM tip has been removed the smooth outer surface. The lateral scale is 840nm. Clearly, a portion of the outer cell wall has been removed exposing the peptidoglycan strands beneath (middle of image). Some of the original surface is visible in the upper left corner, while fragments of the cell wall that have been removed, remain visible within the exposed area as indicated by the arrows.

**Fig.5.**  Some of the strands seem to be twisted pairs as indicated by black arrows.

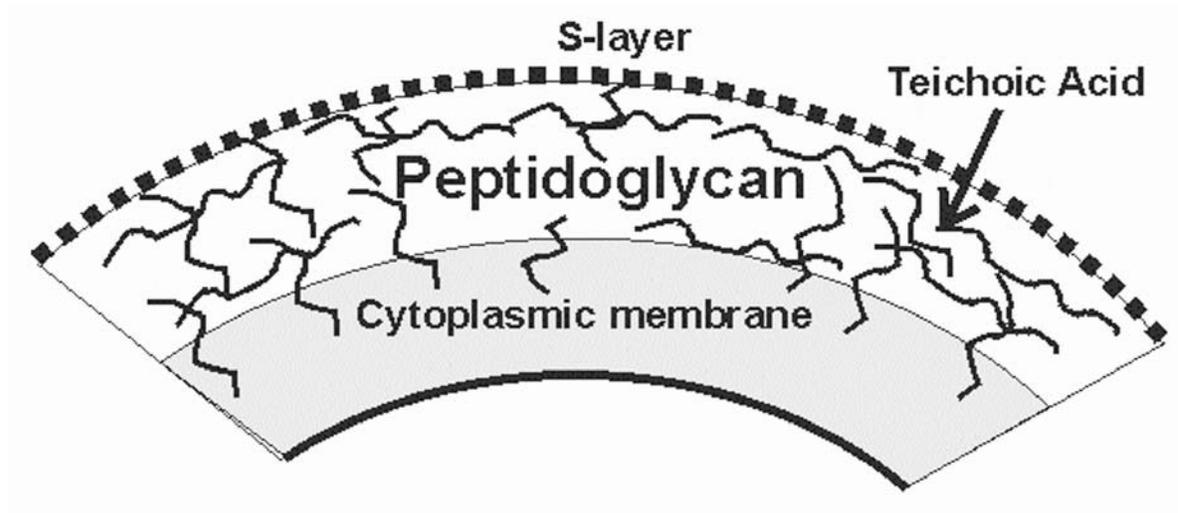

Figure 1 a

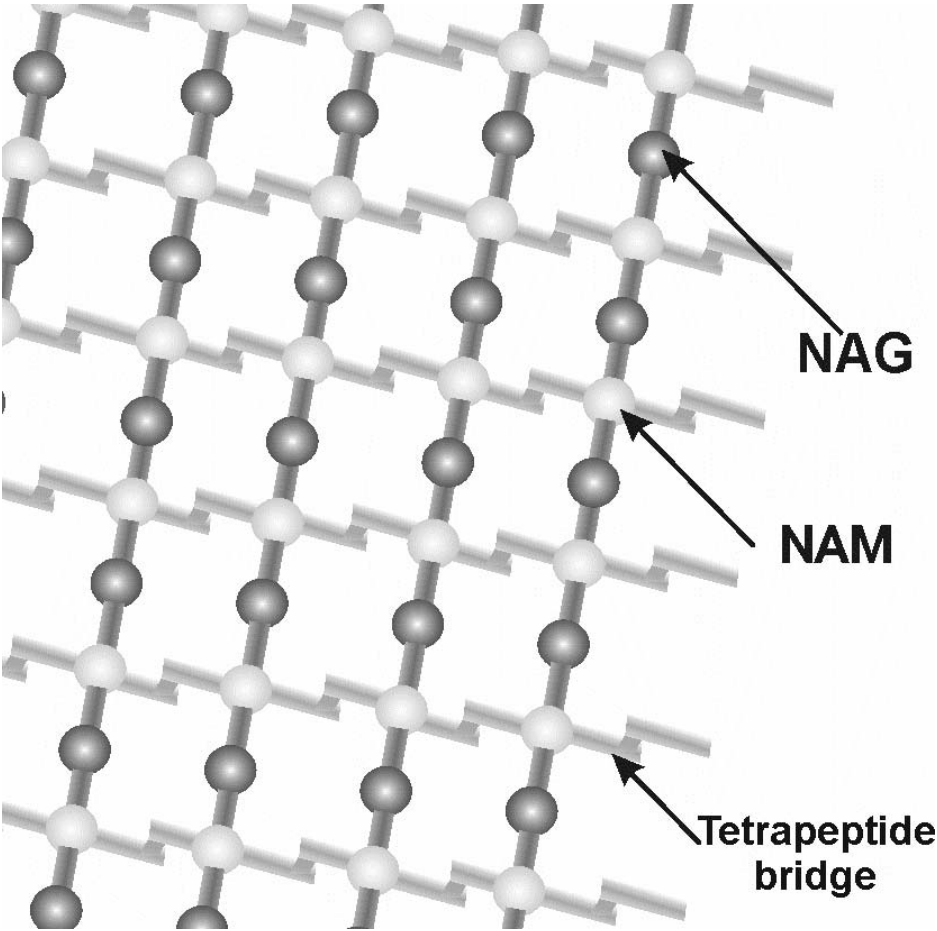

NAG

NAM

Tetrapeptide
bridge

Figure 1 b

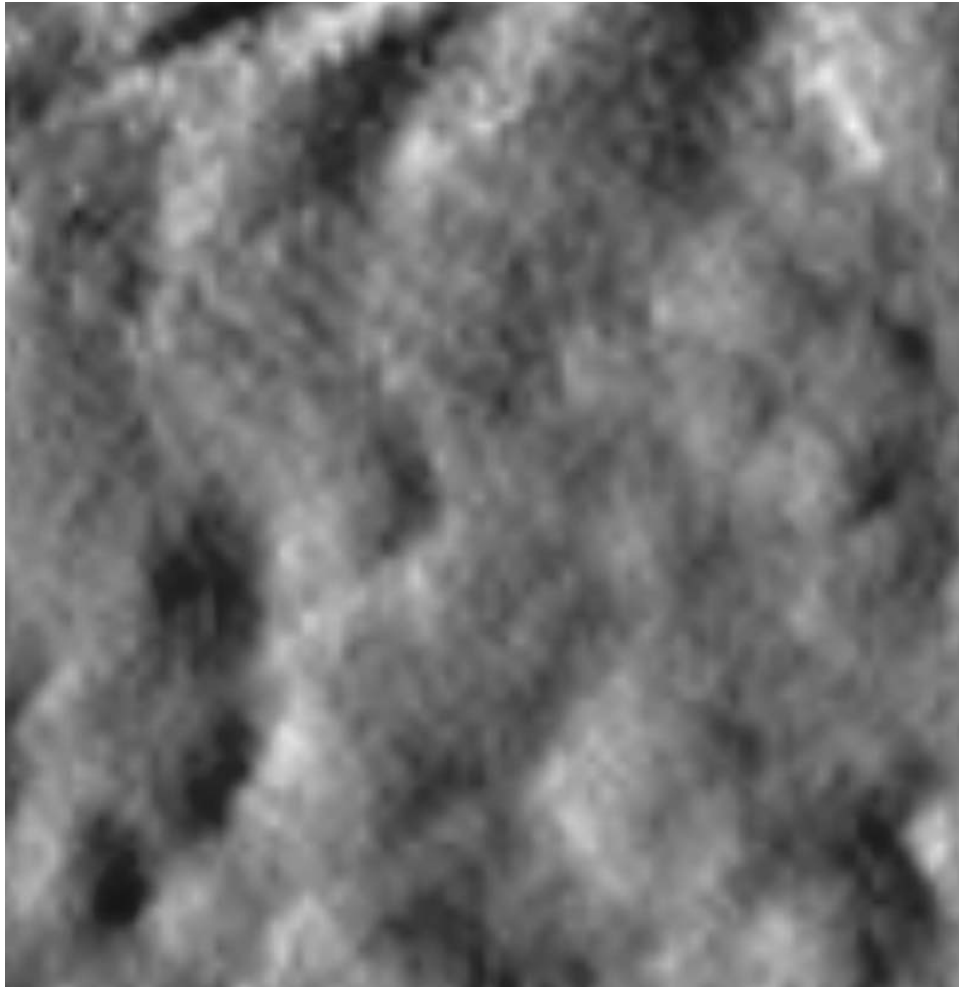

Fig.2

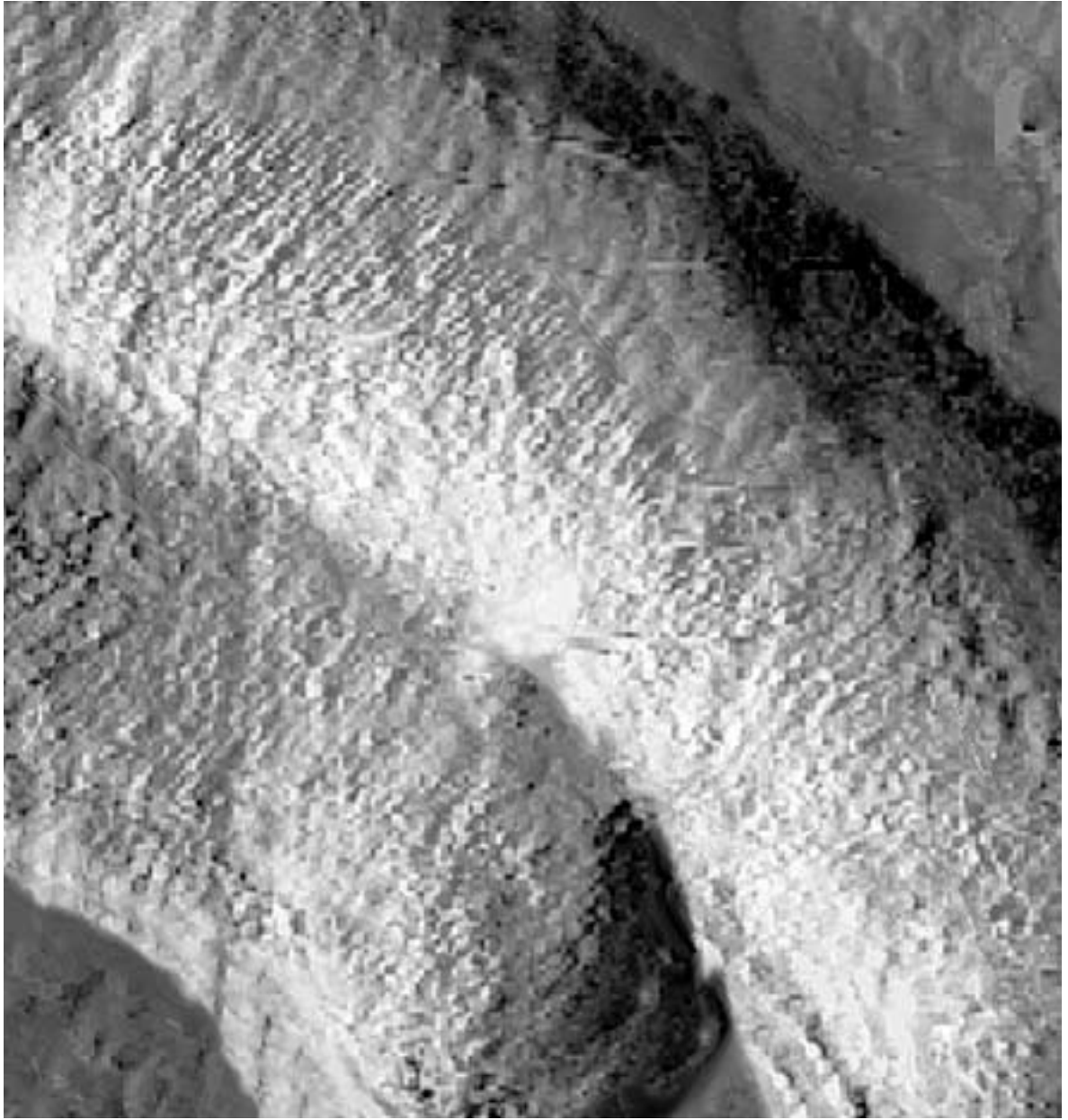

Fig.3, a

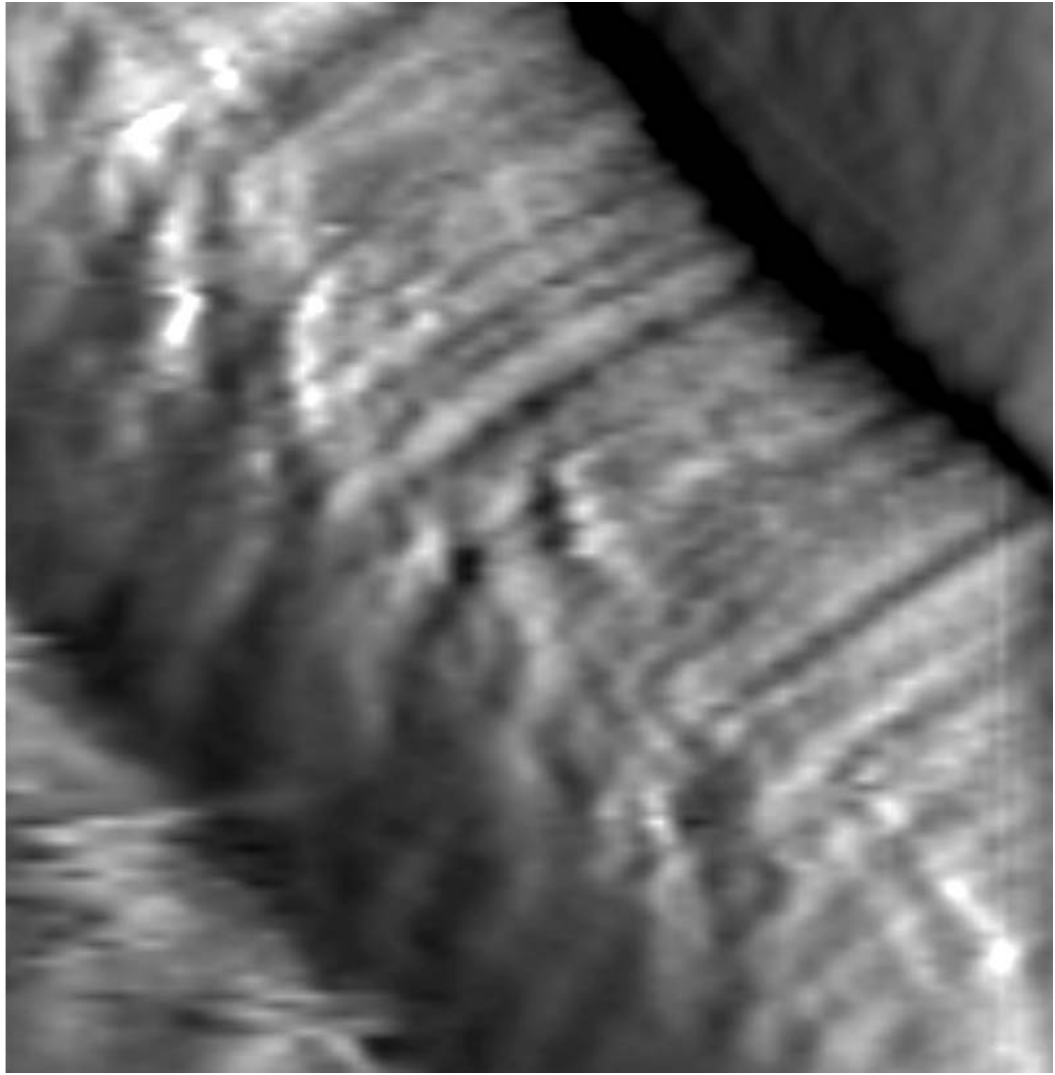

Fig.3 b

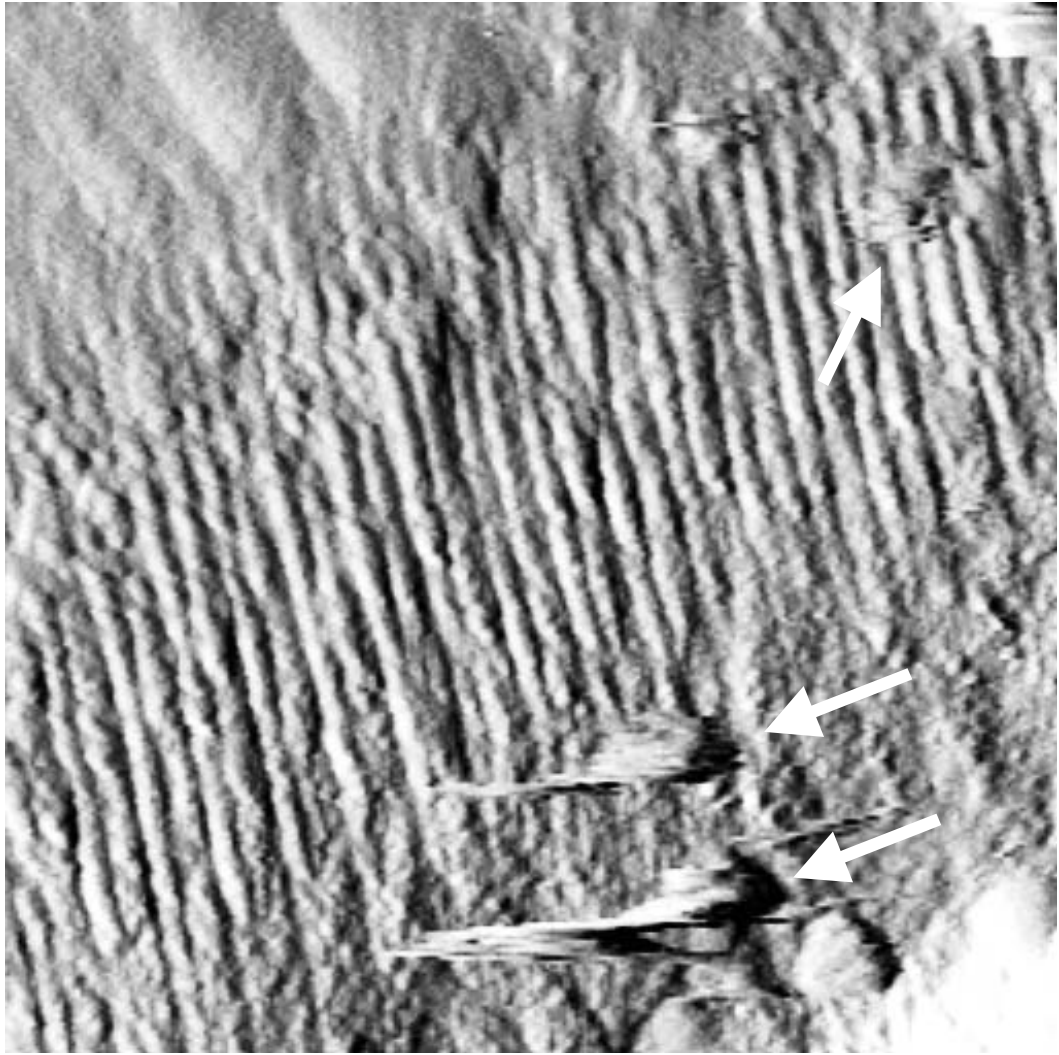

Fig.4

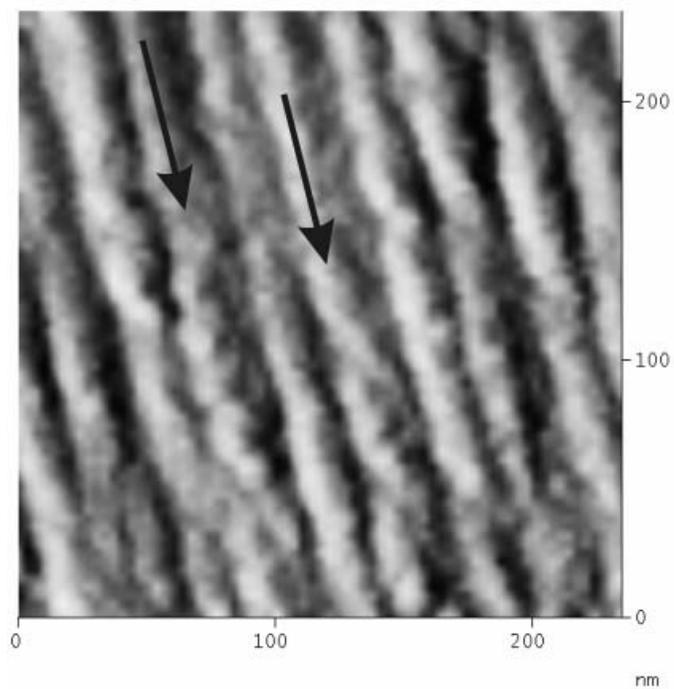

Fig.5